\documentclass[reprint,amsmath,amssymb,aip,twocolumn,superscriptaddress,showpacs]{revtex4-1}
\usepackage{graphicx}
\usepackage{mathrsfs}
\usepackage{bm}
\usepackage{amsmath}
\usepackage{dcolumn}
\usepackage{epstopdf}
\usepackage{dsfont}
\usepackage{amssymb}
\usepackage{tabularx}
\usepackage{array}
\usepackage{float}
\usepackage{color}
\usepackage{epstopdf}
\usepackage{mathrsfs}
\makeatletter
\renewcommand*{\@fnsymbol}[1]{\ensuremath{\ifcase#1\or * \else * \fi}}

\makeatother
\usepackage[colorlinks=true, letterpaper=true, pdfstartview=FitV, linkcolor=blue, citecolor=blue, urlcolor=blue]{hyperref}
\begin{document}
\title{Diverse magnetism in stable and metastable structures of CrTe}

\makeatletter
\renewcommand*{\@fnsymbol}[1]{\ensuremath{\ifcase#1\or *\or \dag\or \dag \or
    \mathsection\or \mathparagraph\or \|\or **\or \dag\dag
    \or \dag\dag \else\@ctrerr\fi}}
\makeatother

\author{Na Kang}
\author{Wenhui Wan}
\email{wwh@ysu.edu.cn}\affiliation{State Key Laboratory of Metastable Materials Science and Technology \& Key Laboratory for Microstructural Material Physics of Hebei Province, School of Science, Yanshan University, Qinhuangdao, 066004, China}
\author{Yanfeng Ge}
\author{Yong Liu}
\email{yongliu@ysu.edu.cn}\affiliation{State Key Laboratory of Metastable Materials Science and Technology \& Key Laboratory for Microstructural Material Physics of Hebei Province, School of Science, Yanshan University, Qinhuangdao, 066004, China}


\begin{abstract}
  In this paper, we systematically investigated the structural and magnetic properties of CrTe by combining particle swarm optimization algorithm and first-principles calculations. With the electronic correlation effect considered, we predicted the ground-state structure of CrTe to be NiAs-type (space group  $P$6$_{3}$/$mmc$) structure at ambient pressure, consistent with the experimental observation. Moreover, we found two extra meta-stable $Cmca$ and $R$$\overline{3}$$m$ structure which have negative formation enthalpy and stable phonon dispersion at ambient pressure. The $Cmca$ structure is a layered antiferromagnetic metal. The cleaved energy of a single layer is 0.464 J/m$^2$, indicating the possible synthesis of CrTe monolayer. $R$$\overline{3}$$m$ structure is a ferromagnetic half-metal. When the pressure was applied, the ground-state structure of CrTe transitioned from $P$6$_{3}$/$mmc$ to $R$$\overline{3}$$m$, then to $Fm$$\overline{3}$$m$ structure at a pressure about 34 and 42 GPa, respectively. We thought these results help to motivate experimental studies the CrTe compounds in the application of spintronics.
  \end{abstract}
\pacs{71.20.Lp, 75.50.-y, 75.50.Ee}

\maketitle

The present spintronic devices mainly consist of ferromagnetic (FM) materials, which have shortcomings such as low integration, slow operating frequency and so on.~\cite{fm} In 2006, antiferromagnetic (AFM) metals are predicted to have giant magnetoresistance,\cite{afm1} which was then observed in expreiment.~\cite{afm2} Antiferromagnets have several advantages in spintronic application, such as high magnetic susceptibility, high switching frequency, the absence of stray fields and strong spin transfer torque capability.~\cite{afma,WANG2017208}
Magnetic transition-metal chalcogenides (TMCs) with rich structural and magnetic properties are regarded as promising candidate materials in spintronics.\cite{TMC1,TMC2,TMC3} It is significant to explore the unknown FM and AFM structures of TMCs.

Here we focus on the CrTe compounds. On the experimental side, according to the phase diagram determined by Ipser et al.,~\cite{Ipser1989} the Cr$_{1-\delta}$Te system has the hexagonal
NiAs (NA) structure for $\delta \leq 0.1$.  NA CrTe is a FM material with a Curie temperature of 340 K.~\cite{2010Tc1}
Furthermore, Wang et al.~\cite{2020two1} synthesized the film of NA CrTe with a thickness of $11\sim45$ nm and found that it still had a hard magnetism.
Apart from NA structure, a meta-stable zinc-blende (ZB) CrTe~\cite{2008ZB} have been grown using molecular-beam epitaxy on GaAs (100) substrate.
T. Eto et al.~\cite{intr2001press} founded that CrTe occurs a structural transition from the NA structure to the MnP (MP) one at pressures between 13 and 14 GPa.
Moreover, a CrTe$_{3}$ compound with a layered structure was synthesized with a long-range AFM order,~\cite{2017CrTe3-2} indicating that layered structure may also exist in the CrTe compounds.

On the theoretical side, using full-potential linearly enhanced plane wave method, Charifi et al.~\cite{compare2018} compared the energy of CrTe in NA, ZB, MP and Rock-salt (RS) structure, and found that NA structure is the ground state. Dijkstra et al.~\cite{first1989} found that NA CrTe has strong Cr 3$d$-Te 5$p$ hybridization and the Cr $3d_{z^2}$-Cr $3d_{z^2}$ overlap along the c axis.
Kanchana et al.~\cite{pressure} predicted that NA CrTe have a magnetic transition from FM to non-magnetic (NM) state in around 45.3 GPa by means of tight-binding linear muffin-tin orbital method.
It was predicted that ZB CrTe have a higher total energy of 0.36 eV compared the NA CrTe.~\cite{ZB2003}
T. Block et al.~\cite{TB} founded that that ZB CrTe is a half-metal with a band gap in the minority channel at Fermi level, due to stronger Cr-Te covalent bonding than NA CrTe.
Liu et al.~\cite{2010Tc2} shown that CrTe in RS structure was only marginally unstable against the NA CrTe, and more stable than the ZB CrTe.
However, Belkadi~\cite{belkadi2018first} recently predicted that the RS structure is the ground-state structure of CrTe by modified Becke-Johnson (MBJ) exchange potential method. Moulkhalwa et al.~\cite{U1} found that the electronic correlation effect has a large influence on structural and magnetic properties of Cr-based chalcogenides, which has not well been considered in previous calculations.

Considering contradictory results in previous works, in this paper, we investigated the crystal and magnetic structures of CrTe using structure searching technology combined with first-principles calculations. With the electronic correlation effect considered, we predicted that the NA structure (space group $P6_{3}$/$mmc$) is the ground-state structure. Two extra meta-stable phases including $Cmca$ and $R$$\overline{3}$$m$ structures are founded. The $Cmca$ phase has a layered structure and is an AFM metal. The $R$$\overline{3}$$m$ phase is a FM half-metal. Their electronic and magnetic properties are carefully analyzed. When the external pressure increases to be above 34 GPa and 42 GPa, the ground-state structure of CrTe changes from $P6_{3}$/$mmc$ to $R$$\overline{3}$$m$, and then to $Fm$$\overline{3}$$m$ structure.

\begin{figure}[t!hp]
\centerline{\includegraphics[width=0.45\textwidth]{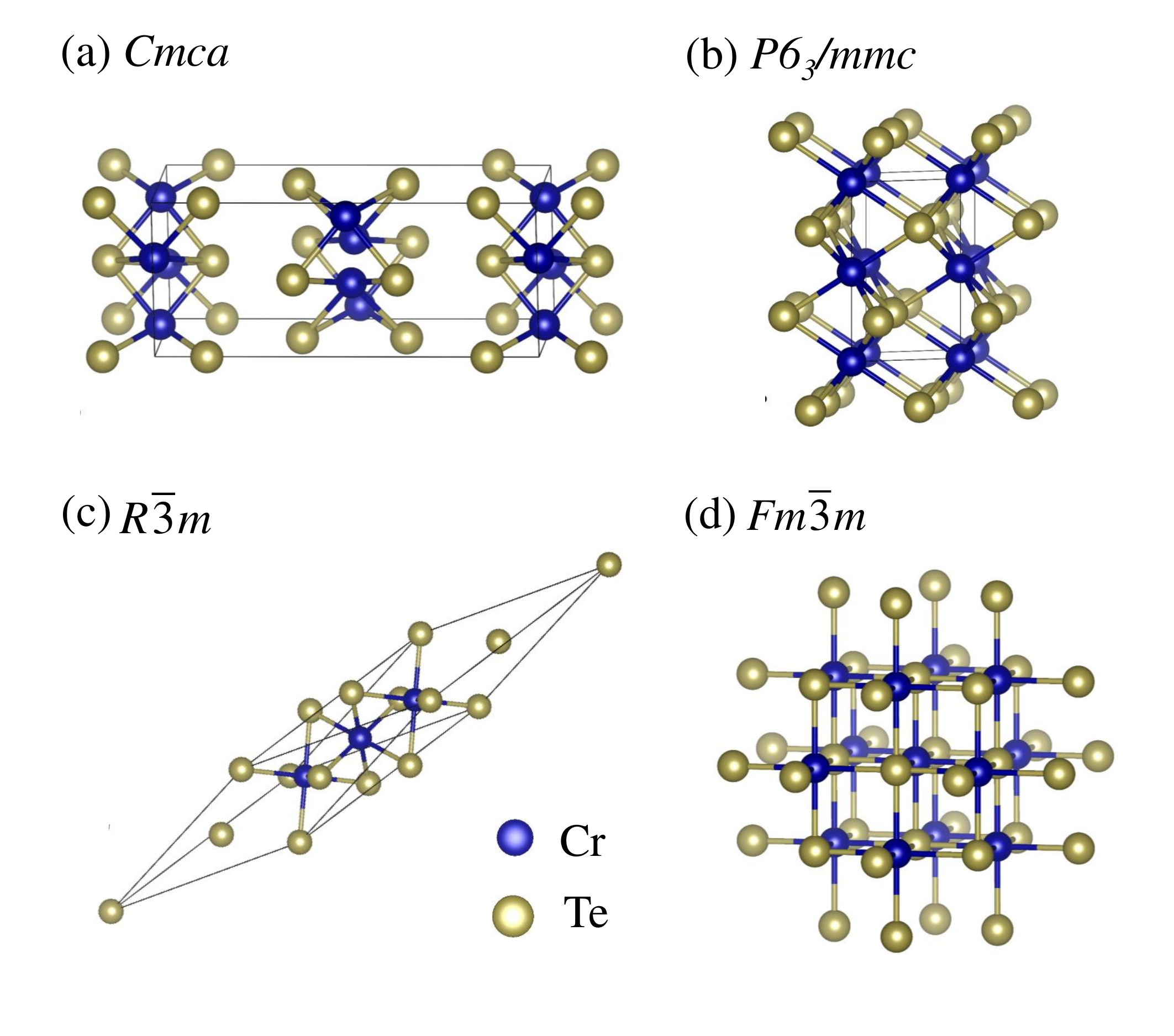}}
\caption{Lattice crystal of predicted CrTe structures including (a) $Cmca$, (b) $P$6$_{3}$/$mmc$, (c)$R$$\overline{3}$$m$ and (d)$Fm$$\overline{3}$$m$ structures.
Blue and brown balls indicate Cr and Te atoms, respectively.
\label{fig:stru}}
\end{figure}

The search of crystal structure was performed by particle swarm optimization algorithm, as implemented in the CALYPSO code.~\cite{caly0} Given the chemical composition and external pressure, the atomic arrangement with the global lowest energy on the potential energy surface can be determined.~\cite{caly0} This algorithm has been successfully applied in various systems with reliability.~\cite{caly1,caly3}
To judge the stability of the CrTe in various structures, we adopted the formation enthalpy ($\Delta\emph{H}$) which is defined as
\begin{equation}
\Delta\emph{H}=[\emph{h}(\rm{CrTe})-\emph{h}(\rm{Cr})-\emph{h}(\rm{Te})]/2,
\end{equation}
where 2 is the number of atoms per CrTe formula. The \emph{h}(CrTe) is the enthalpy per formula corresponding to the most stable magnetic state of CrTe. \emph{h}(Cr) and \emph{h}(Te) are the enthalpy of Cr substance (space group $Im$$\overline{3}$$m$)~\cite{phasecr} and Te substance (space group $P$3$_{1}$21),~\cite{phasete} respectively.
The first-principles calculations were performed using the Vienna ab initio simulation package (VASP).~\cite{vasp1,vasp2}
The projector augmented wave (PAW)~\cite{PAW} pseudopotentials and Perdew, Burke, and Ernzerhof (PBE)~\cite{PBE} exchange correlation functionals are used. A
plane-wave basis set with an energy cutoff of 500 eV and a monkhorst-Pack~\cite{kpoints} k-points with grid spacing of $2\pi \times 0.08 $ \AA$^{-1}$ are used. The crystal structures and atomic positions are fully optimized with a force convergence
threshold of 0.01 eV/\AA. To consider the correlation effects of Cr d orbitals, we applied the PBE + U scheme.~\cite{PBEU} The onsite effective U$_{f}$ value of Cr in CrTe was determined to be 4 eV (see Fig. S1) by the linear response methods.~\cite{effectiveU} The phonon spectrum was calculated using a density functional perturbation theory.~\cite{phonopy1}

\begin{figure}[tbp!]
\centerline{\includegraphics[width=0.45\textwidth]{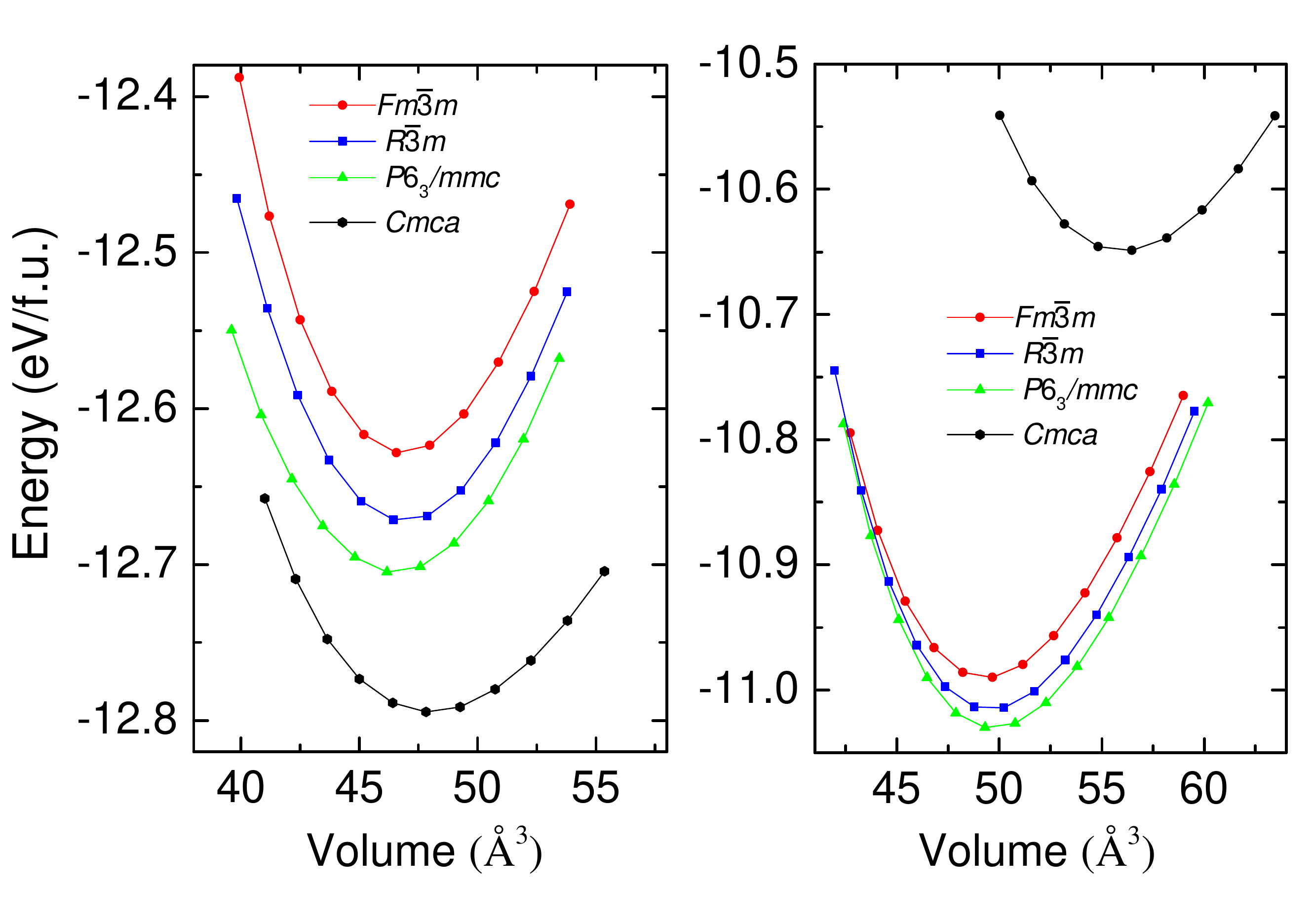}}
\caption{The energy-volume curve for CrTe in four different phases including $Cmca$, $P$6$_{3}$/$mmc$, $R$$\overline{3}$$m$ and $Fm$$\overline{3}$$m$ by (a) PBE and (b) PBE+U methods.
\label{fig:E-V}}
\end{figure}

\begin{table}[!t]\small
\centering \caption{The space group (SG) and group number, the ground magnetic state (MS), lattice parameters(LP), bonding length ($d$$_{Cr-Te}$) and formation enthalpies ($\Delta\emph{H}$ ) of CrTe. The numbers of atoms in a conventional cell are also given.} 
\renewcommand\arraystretch{1.3}  
\begin{tabular}{ccccccccc}
  \hline
  \hline
  SG  &  MS &   LP & $d$$_{Cr-Te}$ & $\Delta\emph{H}$ \\
  atom  &  &  ($\mbox{\AA},\, ^{\circ}$) &  ($\mbox{\AA}$) &   (eV/atom) \\
  \hline
  $Cmca (64)$ & AFM  &$a=5.743$ & 2.814 &  -0.592\\
   16              &&$b=12.671$    &     \\
                   &&$c=6.026$ \\
                   &&$\alpha=\beta=\gamma=90^{\circ}$ \\
  $P6_3/mmc (194)$ & FM   & $a=b=4.216$ & 2.920  & -0.782\\
      4                   &&c = 6.452   &   \\
                          &&$\alpha = \beta = \gamma = 90^{\circ}$ &\\
  $R\overline{3}m$ (166) & FM   & $a=b=c=10.184$ & 2.925  & -0.774\\
      6                         &&$\alpha = \beta = \gamma =23.7^{\circ}\;$ \\

  $Fm\overline{3}m$ (225) & FM   & $a=b=c=5.817$ & 2.909  & -0.762\\
      8                          &&$\alpha = \beta = \gamma =90^{\circ}$ \\
  \hline
  \hline
  \label{table:E}
\end{tabular}
\end{table}

\begin{figure}[tbp!]
\centerline{\includegraphics[width=0.45\textwidth]{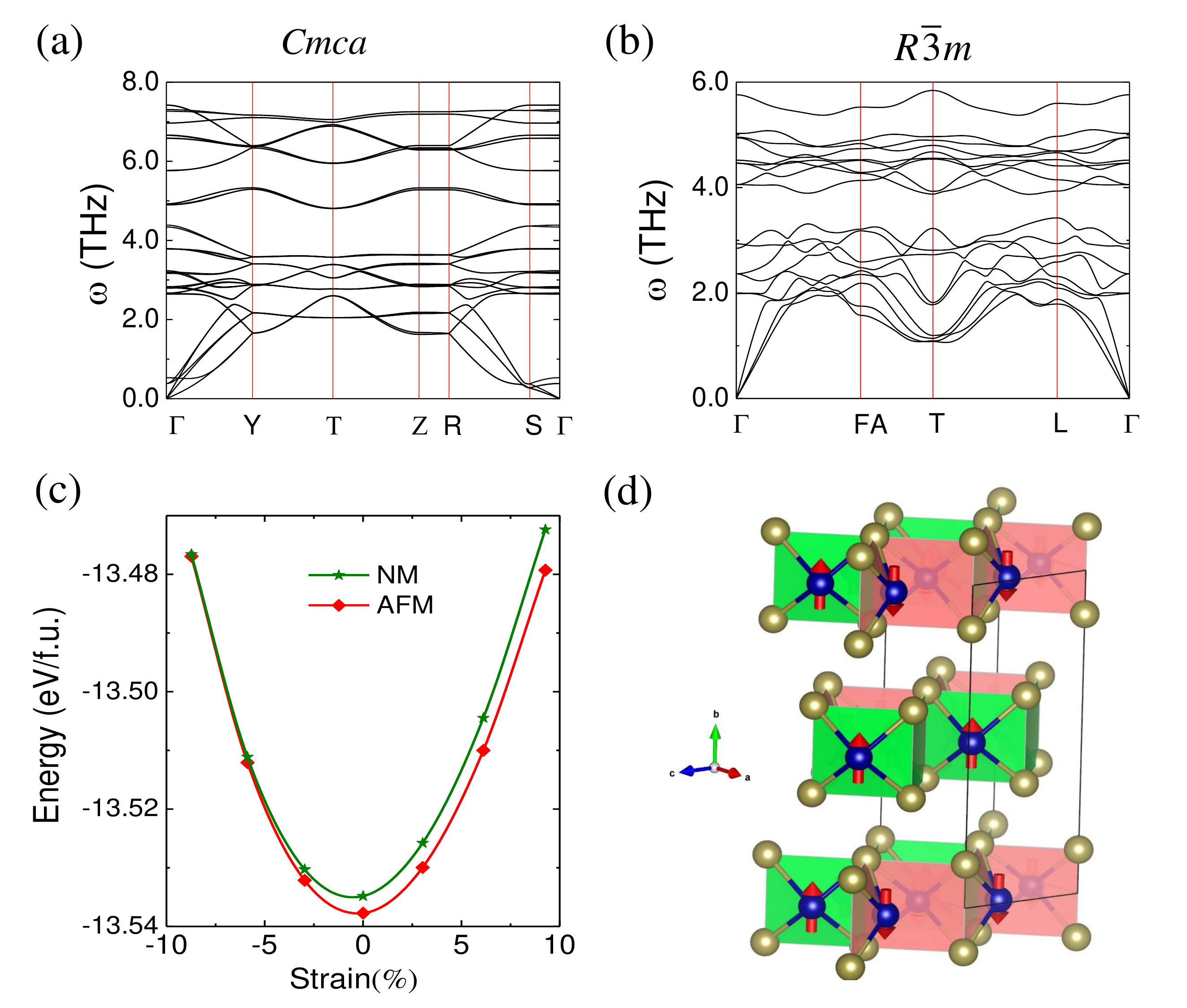}}
\caption{The Phonon spectrum of CrTe in (a) $Cmca$ and (c) $R$$\overline{3}$$m$ phase. (c) The total energy of $Cmca$ phase in NM and the most stable AFM state as function of strain. (d) The spin configuration of $Cmca$ phase in the most stable AFM state. The red arrow indicates the direction of the spin of Cr atoms.
\label{fig:64-E-V}}
\end{figure}

Through the structural search at ambient pressure, several structures with large negative $\Delta\emph{H}$ including $Cmca$, $P$6$_{3}$/$mmc$ (NA), $R$$\overline{3}$$m$ and $Fm$$\overline{3}$$m$ (RS) structures are predicted (see Fig.~\ref{fig:stru}), while the ZB and MnP structures have much smaller negative $\Delta\emph{H}$.
The energy-volume curve of four low-energy structures are displayed in Fig.~\ref{fig:E-V}. Considering the layered structure of $Cmca$ phase, we considered the van der Waals (vdw) functional~\cite{vdw} in the first-principles calculations.
In PBE calculations, the $Cmca$ structure is the ground-state structure and $P$6$_{3}$/$mmc$ structure is meta-stable one (see Fig.~\ref{fig:E-V}(a)). However, in PBE+U calculations, the $P$6$_{3}$/$mmc$ (NA) structure become the ground-state structure (see Fig.~\ref{fig:E-V}(b)) which is consistent with the available  experiment.~\cite{Ipser1989,2010Tc1}
It was indicated that the electronic correlation effect have to be considered in the investigation of the structures of CrTe. Our results are different from previous work,~\cite{belkadi2018first} which adopted MBJ method. However, MBJ method is a potential-only functional and is not suited for computing Hellmann-Feynman forces.~\cite{Choudhary2018} In the following content, we only displayed the results of PBE+U calculations.
The optimized lattice parameters of low-energy CrTe structures are given in Table~\ref{table:E}. The phonon spectrum of four low-energy structures show no imaginary phonon frequencies, confirming their structural stability (see Fig.~\ref{fig:64-E-V}(a, b), and Fig. S2).

Two meta-stable structures including $Cmca$ and $R$$\overline{3}$$m$ structures are predicted, which were not reported in previous works. The $Cmca$ structure has a layered crystal lattice with each layer be held together by vdw interaction. Its single layer consists of Cr$_{4}$Te$_{4}$ cells with four edge-shared CrTe$_{4}$ quadrangular pyramids (see Fig.~\ref{fig:64-E-V}(d))). We have compared the energy of $Cmca$ structure in non-magnetic (NM), FM and different AFM states. It is founded that the FM state was unstable and would change to the NM state. The ground-state magnetic state of the $Cmca$ phase is the AFM state (see Fig.~\ref{fig:64-E-V}(c)), in which the spin distribution in a single layer is opposite (see Fig.~\ref{fig:64-E-V}(d)). The AFM state, which helps to decrease the Pauli repulsion between electrons, results in shorter Cr-Te bond lengths than that of other structures (see Table~\ref{table:E}). Moreover, the calculated cleaved energy of single layer is 0.464 J/m$^{2}$, similar to that of graphene (0.43 J/m$^{2}$)~\cite{225-3}, indicating the possible synthesis of monolayer CrTe.

On the other side, the ground state of $R$$\overline{3}$$m$ structure is the FM state (see Fig. S3). The basic unit of $R$$\overline{3}$$m$ structure is the octahedron, similar to $Fm$$\overline{3}$$m$ and $P$6$_{3}$/$mmc$ structure (see Fig. S4). As seen from Fig.~\ref{fig:E-V}, in both PBE and PBE+U calculations, the energy of $R$$\overline{3}$$m$ structure is slightly higher than that of $P$6$_{3}$/$mmc$ (NA) structure, but lower than $Fm$$\overline{3}$$m$ (RS) structures as well as ZB structures. Thus, it was possible to synthesize the $R$$\overline{3}$$m$ structure in future experiments.

As shown in Fig. S5, the charge density difference of CrTe in various structures indicate that electrons are transferred from Cr atoms to the middle zone of Cr-Te bonds, which indicates a mixture of ionic and covalent bonding interactions between Cr and Te atoms.~\cite{LAPW1} The Cr-Te bond lengths (d$_{Cr-Te}$) of $P$6$_{3}$/$mmc$, $R$$\overline{3}$$m$, and $Fm$$\overline{3}$$m$ structures is longer than that of the $Cmca$ phase (see Table.~\ref{table:E}). The extension of d$_{Cr-Te}$ leads to a decrease in covalent bond interaction, contrary to the ionic bonding interaction.~\cite{2010Tc2} Therefore, the unpaired electrons accumulated on the Cr and Te atoms in CrTe of these three structures slightly increase compared to $Cmca$ structure, leading to a variation of magnetism.

\begin{figure}[tb!]
\centerline{\includegraphics[width=0.45\textwidth]{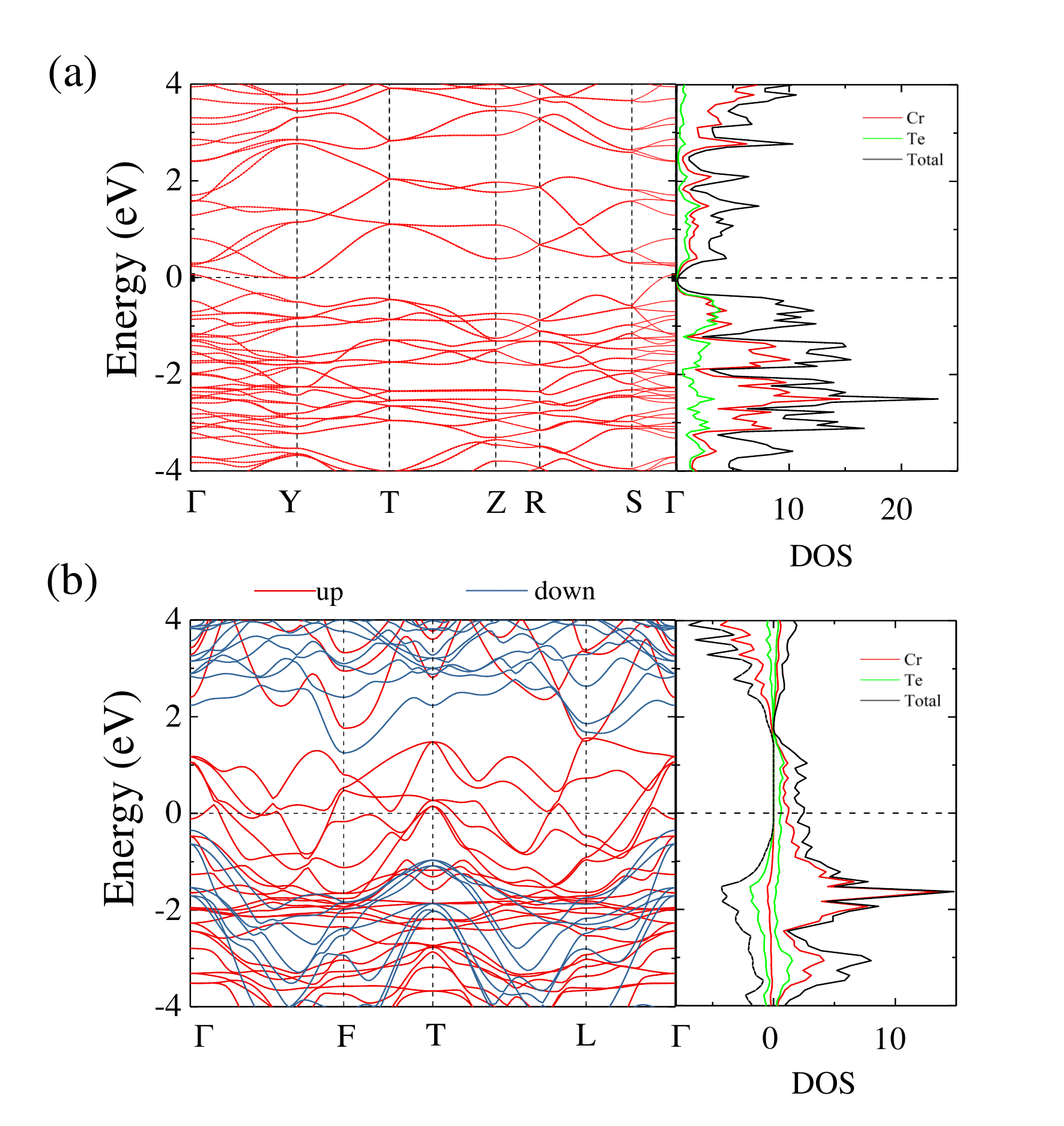}}
\caption{The spin-dependent band structure and partial density of states (DOS) of CrTe in (a) $Cmca$, (b) $R$$\overline{3}$$m$ phases.
\label{fig:dosband}}
\end{figure}

Figure~\ref{fig:dosband} shows spin-dependent band structure and partial density of states (DOS) of the $Cmca$ and $R$$\overline{3}$$m$ structure with U$_{f}$ = 4.0 eV at its equilibrium volume. The $Cmca$ structure is an AFM metal which indicates the possible existence of giant magnetoresistance.\cite{afm1} On the other side, $R$$\overline{3}$$m$ is a FM half-metal with carrier of 100\% spin polarized at Fermi level.
 The DOS near the Fermi level is mainly from the Cr-3$d$ orbitals in both phases (see Fig.~\ref{fig:dosband}).
The analysis of crystal orbital hamilton populations (-COHP) shown that the DOS near the Fermi level consist of anti-bonding states between Cr-3$d$ and Te-5$p$ orbitals (see Fig. S6). Here, we founded that the electron correlation effect has a large influence on the electronic structure of CrTe. For example, the $Cmca$ structure become a AFM semiconductor with a band gap of 0.126 eV if U$_{f}$ of Cr was ignored (see Fig. S7).

\begin{figure}[tb!]
\centerline{\includegraphics[width=0.4\textwidth]{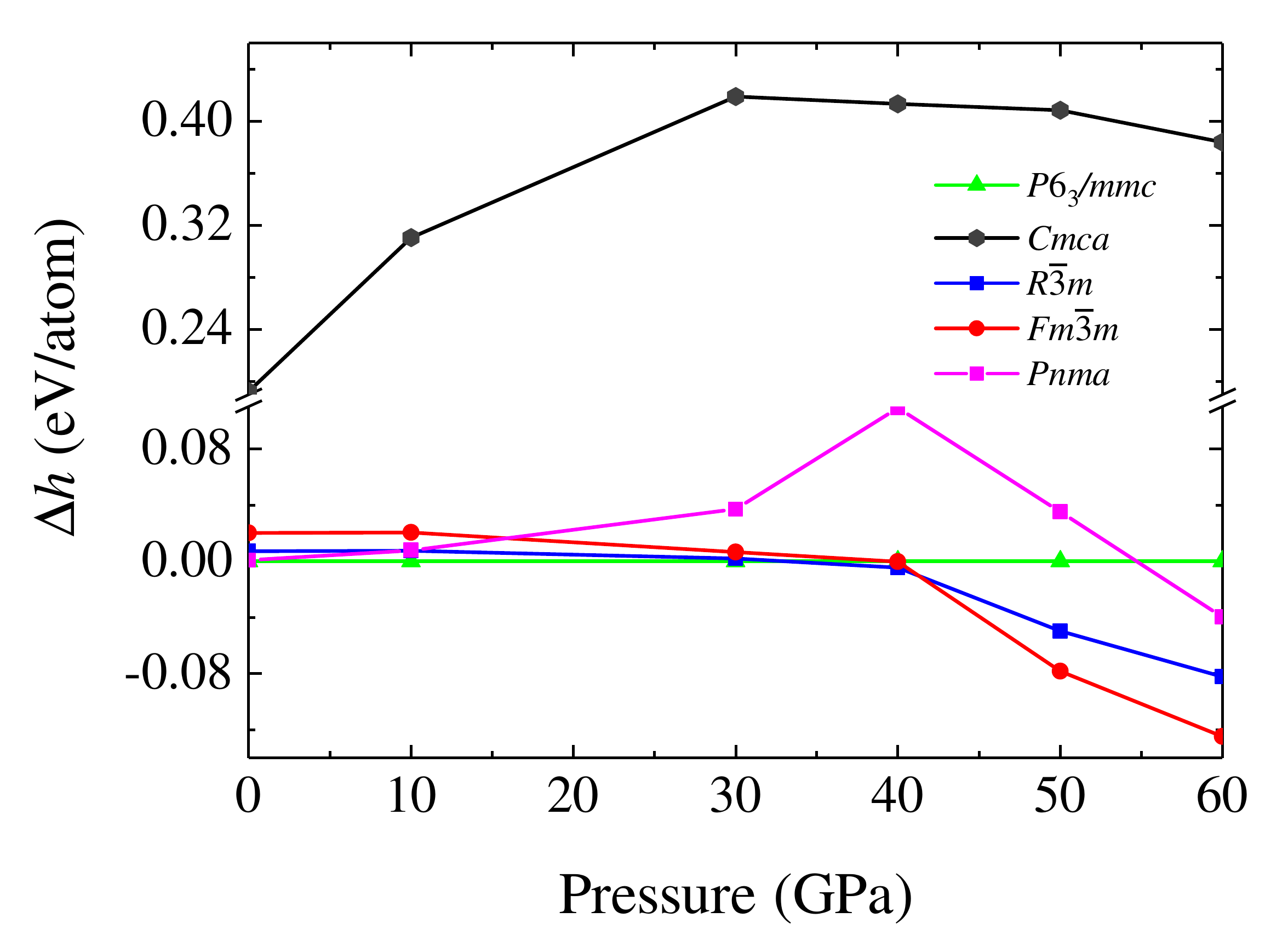}}
\caption{The relative formation enthalpy of $Cmca$, $R$$\overline{3}$$m$, $Fm$$\overline{3}$$m$ and $Pmna$ structure with respect to that of the $P$6$_3$/$mmc$ structure as a function of pressure.
\label{fig:press}}
\end{figure}

We next implored the evolution of the structure of CrTe as a function of external pressure. Figure~\ref{fig:press} plots the pressure-formation enthalpies curve of the CrTe in various structures. We set the formation enthalpies of $P$6$_3$/$mmc$ phase to zero for a comparison. The $P$6$_3$/$mmc$ structure is the most stable structure. However, the energy difference among $P$6$_3$/$mmc$, $R$$\overline{3}$$m$ and $Fm$$\overline{3}$$m$ structures is small. At a range of $30 \sim 45$ GPa, pressure-induced phase transitions occur. The ground-state of CrTe changes from $P$6$_3$/$mmc$ structure to $R$$\overline{3}$$m$ structure and $Fm$$\overline{3}$$m$. The transition critical pressures were estimated as about 34 GPa and 42 GPa by linear interpolation.
During the phase transition, CrTe becomes a FM metal(see Fig. S8). We also check the MnP-type CrTe (space group $Pnma$) which was observed in experiment.~\cite{intr2001press} We predicted that $Pnma$ structure has higher formation enthalpy than $R$$\overline{3}$$m$ and $Fm$$\overline{3}$$m$ structures at high pressure (see Fig.~\ref{fig:press}). Thus, the high-pressure phase of CrTe should be $R$$\overline{3}$$m$ and $Fm$$\overline{3}$$m$ structures. 

In conclusion, we investigated the crystal structures and magnetic properties of CrTe by using particle swarm optimization algorithm combined with the first-principles calculations. It is necessary to consider the electronic correlation effect in determining the ground state of CrTe which is the NiAs-type (space group $P6_{3}/mmc$) structure at ambient pressure, consistent with available experiment. Apart from that, we predicted two meta-stable phases including $Cmca$ and $R$$\overline{3}$$m$ structures. CrTe in $Cmca$ phase is a layered AFM metal. The $R$$\overline{3}$$m$ structures is an FM half-metal and may be synthesized in the future experiment, due to that its energy is slightly higher than ground-state $P6_{3}/mmc$ structure at ambient pressure. The $P$6$_3$/$mmc$ structure is the most stable for low pressure, but the ground-state structure of CrTe would change to $R$$\overline{3}$$m$ and $Fm$$\overline{3}$$m$ structure at the pressure of about 34 and 42 GPa, respectively.

\section*{Supplementary Material}
See the supplemental material for the estimation of  effective onsite U$_{f}$ value; The phonon dispersion of CrTe; The energy-volume curve of CrTe in $R$$\overline{3}$$m$ structures; The charge density difference of CrTe; The crystal orbital hamilton populations (-COHP) of CrTe; The band structure of CrTe in high pressure.

\begin{acknowledgments}
This work was supported by National Natural Science Foundation of China (No. 11904312 and 11904313), the Project of Hebei Educational Department, China (No. ZD2018015 and QN2018012), The Natural Science Foundation of Hebei Province of China (No. A2020203027), and the Doctor Foundation Project of Yanshan University (Grant No. BL19008), and the Scientific Research Foundation of the Higher Education of Hebei Province, China (No. BJ2020015). The numerical calculations in this paper have been done on the supercomputing system in the High Performance Computing Center of Yanshan University.

\end{acknowledgments}


\nocite{*}
\bibliography{CrTe}

\end{document}